# Symmetry Principles and Group Theory in Electromagnetics

Victor Dmitriev

Department of Electrical Engineering, University Federal of Para, PO Box 8619, Agencia UFPA, CEP 66075-900, Belem, Para, Brazil

*Abstract* — **Group theory is used in many textbooks of contemporary physics. However, electromagnetic community often considers group theory as an "exotic" tool. Graduate and postgraduate textbooks on electromagnetics and electrodynamics usually do not contain a description of properties of electromagnetic fields and waves on the basis of group theory. We discuss in this paper from the didactic point of view, several types of symmetry and the corresponding groups and representations. The benefits of the use of group-theoretical methods are a more deep physical understanding of the electromagnetic phenomena and a significant reduction of numerical calculations. These methods are general and exact.**

## I. INTRODUCTION

Symmetry analysis can be effectively used in many parts of electromagnetic theory. It can be applied to differential, integral, variational and matrix description of electromagnetic phenomena. The physical system, coordinate system, classical or quantum-mechanical description are irrelevant.

Symmetry is a transformation that leaves the relevant structure invariant. The collection of symmetries of an object preserving some of its structure forms a group. Group theory can be considered as an instrument to study symmetry and its consequences.

Perhaps, the first mathematicians who introduced the concept of group theory in algebra in context of the solutions of algebraic equations were Evarist Galois and Niels Henrik Abel. At the end of the XIX century, Felix Klein and Sophus Lie developed the theory of discrete groups and continuous groups, respectively. The famous theorem of Emmy Noether proved in 1917 states that every group of symmetry of (differential) equations gives rise to a quantity that is preserved. Introduction of group-theoretical methods to quantum mechanics is connected with the names of H. Weyl, E. P. Wigner, and others. E. P. Wigner received in 1963 the Nobel prize in physics for applying symmetry principles to elementary-particle theory.

About 20-25 years ago, there was a discussion in the European Physical Journal how to teach group theory in the university physical courses~[1]-[2]. Recently, a book for undergraduate level of students was published [3]. Nowadays, many textbooks of contemporary physics contain the group-theoretical description of physical problems (see, for example, the classical 10 volume course of physics of L. D. Landau and E. M. Lifshits).

Group theory is very helpful in many branches of physics: in quantum mechanics, in solid state and molecular physics, in crystallography, and others. A universal classification of differential equations of physics can be based on their symmetry groups [4]. The famous Noether's theorem relates symmetry groups and conservation laws. Notice that even the dimensional analysis of physical quantities can be considered from the group-theoretical point of view [5].

In contrast to the textbooks on physics, the popular graduate and postgraduate textbooks on electromagnetics and electrodynamics for electrical engineering usually do not contain a description of properties of electromagnetic equations, electromagnetic fields, waves and material media on the basis of group theory. For example, in the book [6, p. 267] the author describing Space-Time transformation properties of fields and sources avoids to use group-theoretical language. In the chapter concerning special relativity principles of the textbook [7, p.807], the Lorentz and Poincare groups are only mentioned. However, in some special monographs on electrodynamics, for example, in [8]-[10], the group theory is used extensively.

Sometimes, one can hear an opinion that the group theory and the theory of representations are an "exotic" tool and they are too difficult for students of electrical engineering departments. To our opinion, this theory is not more difficult that the theory of linear spaces (of Hilbert space, in particular) with their 8 axioms [11]. Moreover, these two theories are related. For example, 4 axioms of the 8 of the linear spaces define the Abelian group of eigenvectors (eigenfunctions).

Maxwell's equations possess very high symmetry. All the exact solutions of wave equations in rectangular, circular cylindrical and spherical coordinates follow from symmetry of these differential equations. This symmetry in turn follows from relativistic and gauge invariances of the material properties of propagation medium. Moreover, Maxwell's equations themselves are uniquely determined by their symmetry groups [12].

The orthogonality and completeness properties of sines and cosines, exponentials and special functions such as Bessel, Legendre, spherical harmonics and other classical functions of mathematical physics are defined by symmetry decomposition. A group-theoretical interpretation can be given also to recursion formulas and addition theorems of special functions.

TABLE I. IRREDUCIBLE REPRESENTATIONS OF THE GROUP $C_{4v}$

| $C_{4v}$ | e | $C_2$ | $C_4$ | $C_4^{-1}$ | $\sigma_x$ | $\sigma_y$ | $\sigma_{d1}$ | $\sigma_{d2}$ |
|---|---|---|---|---|---|---|---|---|
| $\Gamma_1$ | 1 | 1 | 1 | 1 | 1 | 1 | 1 | 1 |
| $\Gamma_2$ | 1 | 1 | 1 | 1 | -1 | -1 | -1 | -1 |
| $\Gamma_3$ | 1 | 1 | -1 | -1 | 1 | 1 | -1 | -1 |
| $\Gamma_4$ | 1 | 1 | -1 | -1 | -1 | -1 | 1 | 1 |
| $\Gamma_5$ | $\begin{pmatrix} 1 & 0 \\ 0 & 1 \end{pmatrix}$ | $\begin{pmatrix} -1 & 0 \\ 0 & -1 \end{pmatrix}$ | $\begin{pmatrix} 0 & -1 \\ 1 & 0 \end{pmatrix}$ | $\begin{pmatrix} 0 & 1 \\ -1 & 0 \end{pmatrix}$ | $\begin{pmatrix} 1 & 0 \\ 0 & -1 \end{pmatrix}$ | $\begin{pmatrix} -1 & 0 \\ 0 & 1 \end{pmatrix}$ | $\begin{pmatrix} 0 & -1 \\ -1 & 0 \end{pmatrix}$ | $\begin{pmatrix} 0 & 1 \\ 1 & 0 \end{pmatrix}$ |

The theory of eigenvalues and eigenwaves for symmetrical cases can be easily cast in group-theoretical language. For example, the frequency ω is invariant due to invariance of the medium with respect to translation in Time. The Cartesian components of the wave vector $k_x$, $k_y$ and $k_z$ for an infinite homogeneous medium are the continuous eigenvalues which are invariant due to invariance of the propagation medium with respect to translations in the x-, y- and z-directions, respectively. The TE-TM decomposition of electromagnetic waves in free space and in rectangular waveguides follows from inversion symmetry. The subindexes m and n of the waves $HE_{mn}$ and $EH_{nm}$ for circular dielectric waveguides are also defined by symmetry, and so on.

We believe that basic ideas of the group theory and the theory of representations can be catched with modest efforts even on the undergraduate level. We shall discuss in this paper some basic notions of group theory choosing as examples three important types of symmetry which are met in electromagnetic theory and electrical engineering:
1) discrete rotation-reflexion symmetry (using as an example square microwave waveguide);
2) discrete translational symmetry (considering photonic crystals);
3) Time reversal symmetry.

II. DISCRETE ROTATION-REFLECTION SYMMETRY

As an example of rotation-reflection symmetries met in electromagnetics, let us consider cross-section of a rectangular metallic waveguide shown in Fig. 1a. It is a square possessing the geometrical symmetry $C_{4v}$ (in Schonflies notations [13]). The group $C_{4v}$ contains the following 8 elements of symmetry:
- e is the identity element,
- $C_2$ is a rotation by π around the z-axis,
- $C_4$ and $C_4^{-1}$ are the rotations around the z-axis by π/2 and by -π/2, respectively,
- $\sigma_x$ and $\sigma_y$ are the reflections in the planes x=0 and y=0, respectively,
- $\sigma_{d1}$ and $\sigma_{d2}$ are the reflections in the planes which pass through the axis z and the line a-a and b-b, respectively.

It is easy to demonstrate that all the 4 axioms of the group theory are satisfied for these operations of symmetry. The rotation-reflection operator **R**, which transforms the square into itself, can represent any element of the group.

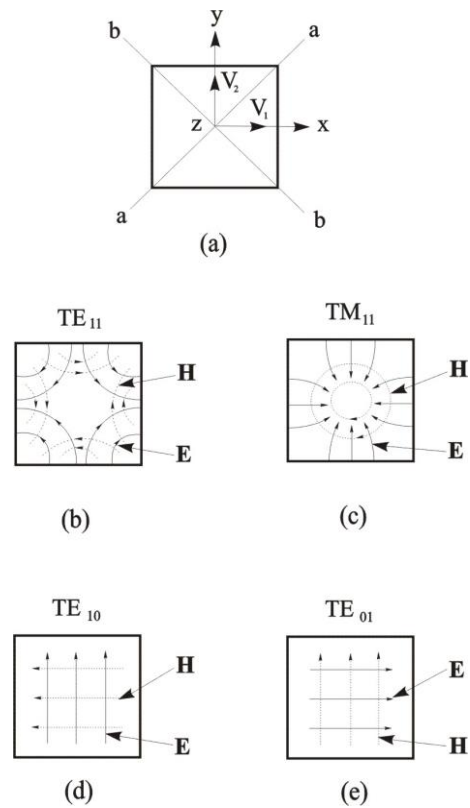

Fig. 1. Cross-section of square metallic waveguide (a), structure of electromagnetic field for the mode $TE_{11}$ (b), for the mode $TM_{11}$ (c), and for the modes $TE_{10}$ (d) and $TE_{01}$ (e).

Now, using this example, we can introduce the idea of classes, representations and irreducible representations of groups. The irreducible representations of the group $C_{4v}$ are given in Table I. The upper line of Table I shows 8 elements of the group. The left column of this Table gives 5 irreducible representations according to 5 classes existing in this group. There are 4 one-dimensional representations $\Gamma_1$, $\Gamma_2$, $\Gamma_3$, $\Gamma_4$, and 1 two-dimensional representation $\Gamma_5$.

Every eigensolution of the boundary-value problem corresponds to one of the irreducible representations, i. e. to one of the rows of Table I. For example, the symmetry properties of the mode $TE_{11}$ (Figure 1b) are completely defined by the representation $\Gamma_3$, i.e. the structure of the electromagnetic field is not changed under the symmetry operations e, $C_2$, $\sigma_x$, $\sigma_y$, and the field changes the sign under $C_4$, $C_4^{-1}$, $\sigma_{d1}$, $\sigma_{d2}$. The mode $TM_{11}$ (Figure 1c) is transformed according to the representation $\Gamma_1$, etc. Here, we can demonstrate also the principal difference in transformation properties of the polar vector *E* and the axial vector *H*.

Now, we can consider the dominant modes $TE_{10}$ (Figure 1d) and $TE_{01}$ (Figure 1e) in a more detail. The electric fields of them have maximum in the planes x=0 and y=0, respectively. In order to discuss the symmetry transformation properties of the electromagnetic fields, we could use the functional space with the eigenfunctions describing these fields. But for the sake of simplicity, we shall consider the corresponding vector space. The modes $TE_{01}$ and $TE_{10}$ are represented by the unit vectors $V_1$ and $V_2$, respectively (Figure 1a). The vector $V_1$ belongs to the row $(\Gamma_5)_{11}$ of the two-dimensional representation $\Gamma_5$, the vector $V_2$ belongs to the row $(\Gamma_5)_{22}$.

In 2D space x0y, the vectors $V_1$ and $V_2$, can be written as follows:

$$V_1 = \begin{pmatrix} 1 \\ 0 \end{pmatrix}, \quad V_2 = \begin{pmatrix} 0 \\ 1 \end{pmatrix}. \quad (1)$$

These two modes $TE_{10}$ and $TE_{01}$ are degenerate because they belong to the same irreducible representation $\Gamma_5$. Under symmetry transformations, they are transformed into each other or into itself. For example, the operator $C_4$ applied to $V_1$ and $V_2$ gives

$$C_4(V_1, V_2) = (V_1, V_2)\begin{pmatrix} 0 & -1 \\ 1 & 0 \end{pmatrix} = (V_2, -V_1), \quad (2)$$

i. e. the basis vector $V_1$ is transformed into $V_2$ and $V_2$ is transformed into $-V_1$. Hence, due to the square form of the waveguide cross-section, here we deal with the two-dimensional polarization degeneracy. Any linear combination of the two degenerate eigenvectors $V_1$ and $V_2$, for example

$$V_1 \pm V_2, \quad V_1 \pm iV_2 \quad (3)$$

is also eigenvector. The combined vectors are o transformed according to the two-dimensional representation $\Gamma_5$.

III. TRANSLATIONAL SYMMETRY

Some examples of electromagnetic devices with translational symmetry are slow-wave structures of traveling wave tubes (TWT) and crossed-field amplifiers, some types of antennas, carbon nanotubes, polymers, photonic crystals, multilayered systems. In order to describe physical systems which are periodical in one directions, one can use line magnetic groups. For 2D problems, diperiodic magnetic groups are used, and for 3D problems-space magnetic groups [13].

Let us consider a simple case of a 1D lattice of periodically spaced dielectric layers with the period d along the axis z (Fig. 2). This lattice can be used for example, as a 1D photonic crystal. The dielectric constant $\varepsilon(z)$ is invariant under translations by any value of $nd$ (*n* is an integer):

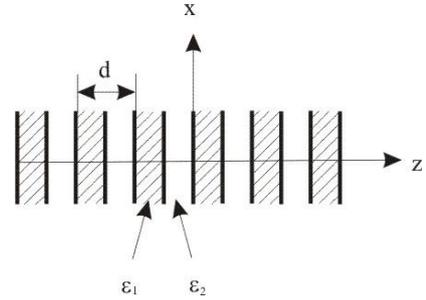

Fig. 1. Cross-section of square metallic waveguide (a), structure of electromagnetic field for the mode $TE_{11}$ (b), for the mode $TM_{11}$ (c), and for the modes $TE_{10}$ (d) and $TE_{01}$ (e).

$$z \rightarrow z' = z + nd. \quad (4)$$

The scalar wave equation for *E* component of the electromagnetic field can be written as follows:

$$OE(z) = \frac{\omega^2}{c^2}E(z), \quad (5)$$

where the operator *O* is

$$O = (\frac{\partial^2}{\partial z^2} - \frac{\varepsilon(z)-1}{c^2}\omega^2). \quad (6)$$

Formally we can introduce an operator *T(n)* which transforms E after any translation defined by (4):

$$E \rightarrow E' = T(n)E. \quad (7)$$

*E* and *E'* must provide equivalent description of the physical system, that is, *E* must be invariant under the transformation *T(n)*. It can be shown that the operator *O* commutes with the operator *T(n)*. It means that the eigenvectors of these two operators *O* and *T(n)* are the same. The operators *T(n)* form a representation of the operator *O*. From (4), we have:

$$T(n)z = z + nd. \quad (8)$$

Using simple geometrical arguments, one can show that the translation operator *T(n)* satisfies all 4 group axioms and, using these axioms, one can find the following relations:

$$E(z+nd) = exp(iknd)E(z), \quad E(z) = V(z)exp(-ikz), \quad (9)$$

where *k* is the wavenumber, *V(z)* is periodic in *z* with period *d*, *exp(iknd)* are the eigenvalues of the problem. Thus, group-theoretical analysis leads to the important result which is known as Bloch theorem. The set of numbers *exp(iknd)* forms a representation of the discrete translation group.

## IV. TIME REVERSAL SYMMETRY

Another important physical symmetry, which is used in classical electrodynamics, is defined by the Time reversal. Literally, the Time reversal operator $\mathcal{T}$ denotes the change of the sign of Time t, i.e. $t \rightarrow -t$. Maxwell's equations are invariant with respect to the Time reversal.

The operator $\mathcal{T}$ as an element of a group and the unit element e form a group where $\mathcal{T}^2$=e. Although for the discrete operator $\sigma_x$, the relation $\sigma_x^2$=e is also fulfilled, there is a difference between $\mathcal{T}$ and $\sigma_x$. The operator $\sigma_x$ is unitary but the operator $\mathcal{T}$ is antiunitary. This leads to some important differences in the results of applications of these two operators.

In our practical application of the operator $\mathcal{T}$, of course, there is nothing of science fictions where one can travel from the present in the past and vice versa. In the magnetic group theory, the operator $\mathcal{T}$ simply reverses direction of motion. In the time domain, as a result, it changes the signs of the quantities which are odd in Time: the velocity, the wave vector, the magnetic field produced by moving charges, the Poynting vector, etc. In the frequency domain, it also complex transposes all quantities [14].

One of the very important consequences of the Time reversal symmetry is the Onsager's theorem, which has a very general nature. Symmetry of the permittivity and permeability tensors with respect to their main diagonals for nonmagnetic media, for example, follows from this theorem. In the theory of microwave circuits, symmetry of scattering matrices for devices with nonmagnetic materials is also a consequence of the Time reversal symmetry.

There are certain difficulties in physical interpretation of the operator $\mathcal{T}$. For example, in the wave equations obtained from Maxwell equations combined with constitutive relations, the Time reversal operator transforms a passive medium in an active one and as a consequence, and a damping electromagnetic wave into a growing one. Thus, the dissipative processes are not Time reversible (notice that in order to overcome this difficulty at least mathematically, Altmann and Suchy [14] suggested to use the so-called restricted Time reversal operator which preserves the passive or active nature of the medium). Another example of these difficulties is as follows: a plane wave diffracted on an object is transformed into a spherical one, but nobody saw the Time reversed process when an incoming spherical wave is transformed into the plane wave (Figure 2). Still another example is transformation of the sources under Time reversal into sinks.

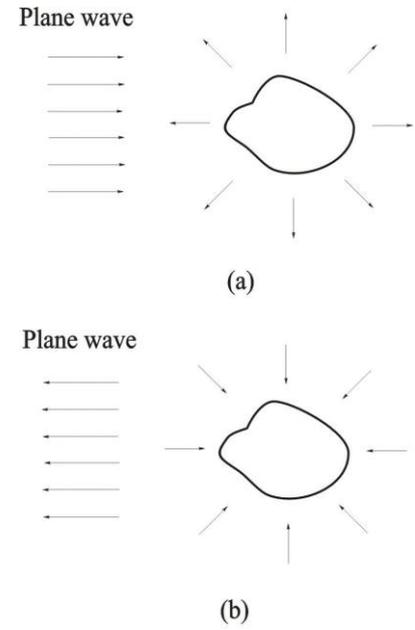

Fig. 2. Scattering of electromagnetic waves in the direct (a) and the Time-reversed (b) processes.

These examples show that the approach described above where the present and the past are reversible, physically is not quite correct. All the real physical processes are not reversible in Time. There exists "the arrow of Time" [15]. In mathematical description of thermodynamical processes, in particular, one of the consequences of this irreversibility is disintegration of the dynamical group of evolution with an operator $U(t)$ into 2 semigroups, one for for $U(+t)$, and the other one for $U(-t)$.

In spite of these difficulties in interpretation of the Time reversal, the idea of use of this operator in classical electrodynamics is very fruitful. In particular, it leads to Lorentz reciprocity theorem [14]. The operator $\mathcal{T}$ is especially useful in the problems involving magnetic media. Many examples of application of the Time reversal operator to electromagnetic problems can be found in [9,14].

Notice that the discussed in Section 2 geometrical group $C_{4v}$, for example, does not describe completely the physical properties of the rectangular waveguide. The full symmetry of the waveguide must include also the Time reversal operator $\mathcal{T}$ and the product of $\mathcal{T}$ with all the geometrical elements of the group $C_{4v}$, i.e. the full magnetic group can be denoted as $C_{4v} + \mathcal{T}C_{4v}$. Notice also, that all the above discrete symmetries with the operators $R$, $T(n)$ and $\mathcal{T}$ can be combined, and the corresponding groups are also useful in some electromagnetic problems, for example, in waveguides with magnetized ferrite elements.

## V. CONCLUSIONS

Group theory is a systematic, concise and very powerful approach to problems, which have symmetry. It provides deep and broad understanding of the problems. The results obtained by group-theoretical methods are general and exact.

Group theory also allows one to reduce considerably the burden of numerical calculations.

Is there any necessity to include the group theory in the electrical engineering courses? We believe, yes. In modern technology, there is a tendency of using more and more complex materials and structures and more and more sophisticated mathematical tools for their electromagnetic analysis and synthesis. The group theory as a universal tool and a unified approach to many problems can render an invaluable help.

A growing number of scientific papers and books on the electromagnetic problems where the authors used group theory is also a justification of our statement. This concerns, for example, scattering and inverse scattering problems (in particular, in target identification) [16]-[18], near-field measurements [19,20]), radar polarimetry [21], electromagnetic and geophysical remote sensing [22], problems of reflection of light in optics [23], waveguiding problems [9], analysis and synthesis of electrical networks [24,25], antennas and antenna arrays [10,26], nanoantennas [27], signal processing [28], problems of propagation of electromagnetic waves in complex media [9,14], Green's tensor calculations [29], for analysis of dielectric arrays [30], and others.

Victor Dmitriev, victor@ufpa.br, Tel +91-3201-7302, Fax +91-3201-7634. This work was supported by Brazilian agency CNPq.